# Coronary Artery Disease Diagnosis; Ranking the Significant Features Using Random Trees Model

**Javad Hassannataj Joloudari [1], Edris Hassannataj Joloudari [2], Hamid Saadatfar [1], Mohammad GhasemiGol [1], Seyyed Mohammad Razavi [3], Amir Mosavi [4,5,6], Narjes Nabipour [7], Shahaboddin Shamshirband [8,9,*], Laszlo Nadai [6]**

[1] Department of Computer Science and Engineering, University of Birjand, Iran; Javad.hassannataj@birjand.ac.ir, saadatfar@birjand.ac.ir, ghasemigol@birjand.ac.ir
[2] Department of Nursing, School of Nursing and Allied Medical Sciences, Maragheh Faculty of Medical Sciences, Maragheh, Iran;
[3] Department of Electronics, Faculty of Electrical and Computer Engineering, University of Birjand, Birjand, Iran; smrazavi@birjand.ac.ir
[4] Faculty of Health, Queensland University of Technology, 130 Victoria Park Road, Queensland 4059, Australia
[5] Institute of Structural Mechanics, Bauhaus Universität-Weimar, D-99423 Weimar, Germany
[6] Kalman Kando Faculty of Electrical Engineering, Obuda University, 1034 Budapest, Hungary
[7] Department Institute of Research and Development, Duy Tan University, Da Nang 550000, Vietnam,
[8] Department for Management of Science and Technology Development, Ton Duc Thang University, Ho Chi Minh City, Vietnam
[9] Faculty of Information Technology, Ton Duc Thang University, Ho Chi Minh City, Vietnam
* Correspondence: shahaboddin.shamshirband@tdt.edu.vn

**Abstract:** Heart disease is one of the most common diseases in middle-aged citizens. Among the vast number of heart diseases, the coronary artery disease (CAD) is considered as a common cardiovascular disease with a high death rate. The most popular tool for diagnosing CAD is the use of medical imaging, e.g., angiography. However, angiography is known for being costly and also associated with a number of side effects. Hence, the purpose of this study is to increase the accuracy of coronary heart disease diagnosis through selecting significant predictive features in order of their ranking. In this study, we propose an integrated method using machine learning. The machine learning methods of random trees (RTs), decision tree of C5.0, support vector machine (SVM), decision tree of Chi-squared automatic interaction detection (CHAID) are used in this study. The proposed method shows promising results and the study confirms that RTs model outperforms other models.

---

## 1. Introduction

Today, we face a huge amount of data in industry as well as organizations such as the Healthcare [1,4]. Since data collection and analysis are difficult, time consuming and costly, we are always looking for a way to optimum use of data to achieve the correct decision that can be referred to diagnose and experiment of diseases in healthcare organizations [3]. In addition, common method such as angiography [5,6] in experimenting and diagnosing diseases is costly and have adverse effects for patients as healthcare researchers are trying to utilize methods that avoid the high cost as well as the adverse effects of previous methods, which can be performed by using computer-aided disease diagnose methods means machine learning. Whereas, data mining process by utilizing machine learning science and database management knowledge [1] has become a robust tool for data analysis and management of health industry data which ultimately leads to knowledge extraction.

It should be noted that, with the progress of technology in the healthcare especially, healthcare industry 4.0, human lifetime has become progressive and more comfortable [7]. In this new generation, with the development of new medical devices, equipment and tools, new knowledge can be gained in the field of disease diagnosis. One of the best ways to quickly diagnose diseases is to use computer-assisted decision making, i.e. machine learning to extract knowledge from data. In general, knowledge extraction from data is an approach that can be very crucial for the medical industry in diagnosing and predicting diseases. In other words, the purpose of knowledge extraction is the discovery of knowledge from databases in the data mining process. Data mining is used as a suitable approach to reduce costs and quick diagnosis of the disease.

Therefore, the purpose of the data mining process, known as database knowledge discovery (KDD), is to find a suitable pattern or model of data that was previously unknown so that these models can be used for specific disease diagnosis decisions in the healthcare environment [1]. Steps to the KDD process [1] include data cleaning (to remove disturbed data and conflicting data) and data integration (which may combine multiple data sources), data selection (where appropriate data is retrieved from the database for analysis), and Data transformation (where data are synchronized by performing summary or aggregation operations and transformed appropriately for exploring), data mining (the essential process in which intelligent methods are used to extract data patterns), pattern evaluation (to identify suitable patterns that represent knowledge based on fit measurement's) and knowledge presentation (where visualization and presentation techniques are used to provide users with explored knowledge) are shown in Figure. 1.

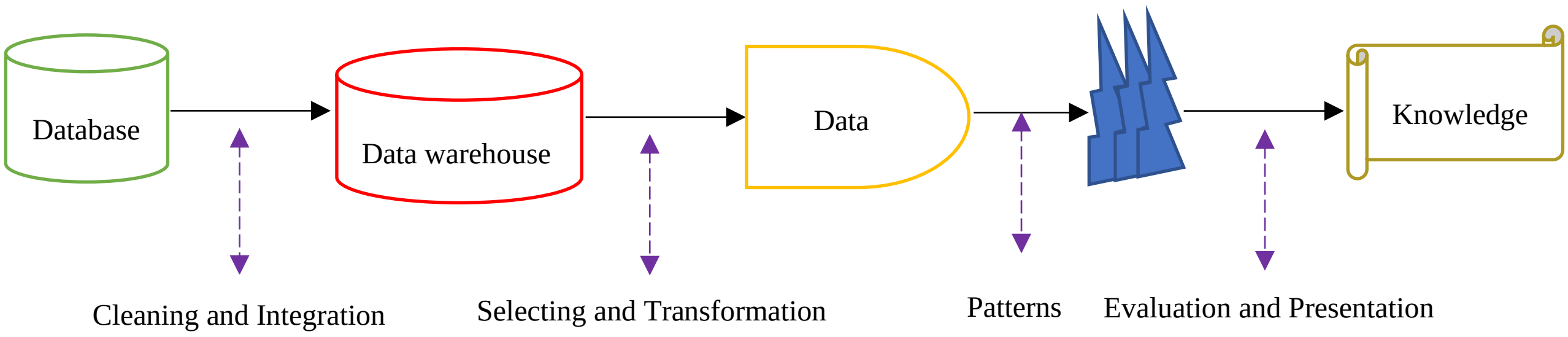

**Fig. 1.** KDD process steps [1].

Given that the subject of this study is in the field of heart disease. Heart disease encompasses a variety of conditions, including congenital diseases, coronary artery disease, and heart rheumatism. Among these conditions, coronary artery disease is the most common so that comprehensive reports of heart disease have been conducted in recent years on heart disease, it should be said that, the World Health Organization (WHO) has declared that coronary artery disease (CAD) as the most common type of cardiovascular disease is a crucial cause of death worldwide [8]. More than 30% of deaths worldwide were due to CAD, which resulted in more than 17 million deaths in 2015 [9]. Also, more than 360,000 Americans have died from heart attacks [10]. As a result, heart disease costs alone total more than $200 billion in the United States annually [10]. In addition, health care costs for heart disease will double by 2030, according to the American Heart Association [11].

Hence, in this paper, has been paid attention on heart disease case study in order to apply a prediction method on coronary artery disease [6,12]. One way to accurately diagnose this disease is to use data mining methods to build an appropriate and robust model that is more reliable than medical imaging tools, including angiography in the field of diagnosis of coronary heart disease [4-6]. A main challenge in model learning is the feature selection problem so that the feature selection step is so important in data mining and its purpose is to eliminate unnecessary and unimportant features [1, 13-15]. The method used in this study is feature ranking-selection method to choose the best subset of features in dataset. In this method, we utilize various data mining methods including Random trees (RTs), decision tree of C5.0, Chi-Squared Automatic Interaction Detection (CHAID) and support vector machine (SVM) that through these methods, select the subset of features according to their order of priority, takes place. For this purpose, the subset of features is ranked from the least important to the most important due to the different weightings to the features associated with the classification models that these features were assigned in output simulator.

Finally, among the classification models used in this study, obtaining the most appropriate subset feature by Random trees model with the best classification set and the most accurate classification of coronary-heart disease diagnosis is the main purpose of this study. As a result, in terms of accuracy, area under the curve (AUC) and Gini value criteria for CAD diagnosis, Random trees model is the best model compared to other prediction models.

The rest of the paper is organized as follows: The present study is expressed data mining classification methods in Section 2 and related works is described in section 3. The proposed methodology is explained in Section 4. Section 5 represents the evaluated results of the experiment. Sections 6 presents findings of the research and the conclusions, namely "Results and Discussion" and "Conclusion and Future works" in section 7.

## 2. Data Mining Classification Methods

In this section, we describe the classification methods used in this study. These methods include CHAID decision tree, C5.0 decision tree, Random trees and support vector machine (SVM). Among the mentioned methods, except for the support vector machine, CHAID, C5.0, Random trees (RTs) because are based on the decision tree, rules are extracted that are useful for the diagnosis of CAD especially rule extraction using RTs.

### *2.1. Decision tree of CHAID*

The Chi-Squared Automatic Interaction Detection (CHAID) is one of the oldest tree classification, and it is a supervised learning methods by building the decision tree, which is evidence of the rules extraction, which is proposed by Kass [16]. This classification model is a statistical method based on the diagnosis of Chi-squared automatic interaction, and it is a recessive partitioning method that can be given by input features as predictors and the predictive class, a Chi-squared statistic test between target class and the Predictive each feature are computed [17-19] so that the predictive features are ranked in order of their priority. As such, the most significant predictors of subset feature with the highest probability of their weight to diagnose CAD to be gained. It should be noted that the process of selecting a significant predictor feature is based on data sample segmentation so that until we reach an external node i.e. the leaf, the samples partition continues into smaller subdivisions [17,20].

In general, the CHAID model includes the following steps [17-19]:

1. *Reading predictors*; The first step is to make classified predictors or features out of any consecutive predictors by partitioning the concerned consecutive disseminations into a number of classifies with almost equal number of observations. For classified predictors, the classifies or target classes are determined.
2. *Consolidating classifies*; The second step is to round through the features to estimate for each feature the pair of features classifies that is least significantly different with concern to the dependent variable. In this process, the CHIAD model includes two types of statistical tests. One, for classification dataset, it will gain a Chi-square test or Pearson Chi-square. The assumptions for Chi-square test are as follows:
   $N_{ij}$ = The value of observations concerned with feature fields or sample size,
   $G_{ij}$ = The gained expected feature fields for datasets, for example, the training dataset $(xn = i, yn = j)$,
   $V_n$ = The value weight ($W_n$) concerned with per sample of dataset,
   $D_f$ = The most number of logically independent values, which are values that have the freedom to vary, in the dataset, namely, Degrees of Freedom. $D_f$ is equal to ($Nij$-1).
   $C$ = The corresponsive data sample, afterward:

$$X^2 = \sum_{j=1}^{J} \sum_{i=1}^{D} \frac{(N_{ij}-G_{ij})^2}{G_{ij}} \quad (1)$$

$$N_{ij} = \sum_{N \epsilon C} F_n D_f \, (X_n = i \, \cap \, Y_n = j) \quad (2)$$

Two, for regression datasets where the dependent variable is consecutive, in other words, for variables based on measure-dependent, $F$-tests. If the concerned test for a given pair of feature classifies is not statistically significant as defined by an alpha-to-consolidate value, then it will consolidate the concerned feature classifies and iterate this step, i.e., obtain the next pair of classifies, which now may include previously consolidated classifies. If the statistical significance for the concerned pair of feature classifies is significant, i.e., less than the concerned alpha-to-consolidate value), then it will gain optionally a Bonferroni adopted $p$-value for the set of classifies for the concerned feature.

$$F = \frac{\sum_{i=1}^{D} \sum_{N \epsilon C} W_n V_n D_f (X_n = i) \left(Y_i' - Y\right)^2 / (D_f - 1)}{\sum_{i=1}^{D} \sum_{N \epsilon C} W_n V_n D_f (X_n = i) (Y_n - Y')^2 / (N_f - D_f)} \quad (3)$$

given that the functions $Y_n$, $Y$, and $N_f$ are formulated as follows:

$$Y_n = \frac{\sum_{N \epsilon C} W_n V_n Y_n D_f (X_n = i)}{\sum_{N \epsilon C} W_n V_n D_f (X_n = i)} \quad (4)$$

$$Y = \frac{\sum_{N \epsilon C} W_n V_n Y_n D_f}{\sum_{N \epsilon C} W_n V_n D_f} \quad (5)$$

$$N_f = \sum_{N \epsilon C} V_n \quad (6)$$

3. *Selecting the partition variable*; The third step is to select the partition the predictor variable with the smallest adapted $p$-value, i.e., the predictor variable that will gain the most significant partition. The $P$-value is formulated in a $(P = pr(X_c^e > X^2))$. If the smallest (Bonferroni) adopted p-value for any predictor feature is greater than some alpha-to-partition value, then no further partitions will be done, and the concerned node is a final node. Continue this process until no further partitions can be done, i.e., given the alpha-to-consolidate and alpha-to-partition values). Eventually, according to step 2, the p-value is obtained as follows:

$$P = P(F(C - 1, N_f - 1) > F) \quad (7)$$

*2.2. Decision tree of C5.0*

Following is the process of improving decision tree models including ID3 [21,22], C4.5 [23-25], the C5.0 tree model [26-29] as the latest version of decision tree models developed by Ross. The improved C5.0 decision tree is manifold faster than its ally models in terms of speed. This model in terms of memory usage, the memory gain is much higher than the other models mentioned. The model also improves trees by supporting boosting and bagging [25] so that using it increases accuracy of diagnosis. As one of the common characteristics among decision trees is weighting to disease features, but the C5.0 model allows different features and types of incorrect classifies to be weighted.

One of the crucial advantage of the C5.0 model to test the features is gain ratio which the information gain is increase, i.e., the information entropy, and the bias is reduce [1,17,29]. For example, the assumptions for The information entropy, information gain, and gain ratio problems are as follows [1,17,25]:

we assume the $S$ as a set of training dataset and splits S into n subsets, and, $N_i$ = The sample dataset of $K$ features.

So, we obtain the features to diagnosis CAD selected with the least information entropy, and the most information gain and gain ratio. The information entropy, information gain, and gain ratio are formulated as follows.

$$Info_{\,Gain}(S, K) \; = \; \sum_{i=1}^{N \in C_i} P_i \, \times Info_{\,Entropy}(S_i) \quad (8)$$

$$Info_{\,Entropy}(S) \; = \; - \sum_{j=1}^{N \in C_i} P_i \, log_2 \, P_i \quad (9)$$

Based on formulas (8) and (9), the number of *K* features, a partition *S* according to values of *K*, and where *P* is the probability distribution of division (*C1*, *C2*, …, *Ci*):

$$P = (|C1|/|S|, |C2|/|S|, \dots, |Ci|/|S|) \quad (10)$$

Based on formula (10), where *Ci* is the number of disjoint classes and $|S|$ is the number of samples in set of S. The value of Gain is computed as follows.

$$Gain\,(S,K) = Info_{\,Entropy}(S) - Info_{\,Gain}(S,K) \quad (11)$$

Ratio instead of Gain was suggested by Quinlan so that *Split Info (K,C)* is the information due to the division of *C* on the basis of value of categorical feature *K*, using the following:

$$Split\ Info(K,C) = Info_{\,entropy}(|C1|/|C|, |C2|/|C|, .., |Ci|/|C|) \quad (12)$$

$$GainRatio(K,C) = Gain(K,C)\ /\ Split\ Info(K,C) \quad (13)$$

For formulas of (12) and (13), where $(C1, C2, \dots Ci)$ is the partition of $C$ induced by value of $K$.

### *2.3. Support Vector Machine*

Support Vector Machine (SVM) is a supervised learning model based on statistical learning theory and structural risk minimization [29,30] presented by Vapnic as only the data assigned in the support vectors are based on machine learning and model building. The SVM model is not sensitive to other data points and its aim is to find the best separation line, i.e. the optimal hyperplane between the two classes of samples so that it has the maximum distance possible to all the two classes of support vectors [29-32]. The predictor feature is determined by the separator line for each predictive class. Fig. 2 shows the scheme of the support vector machine in 2-dimensional space.

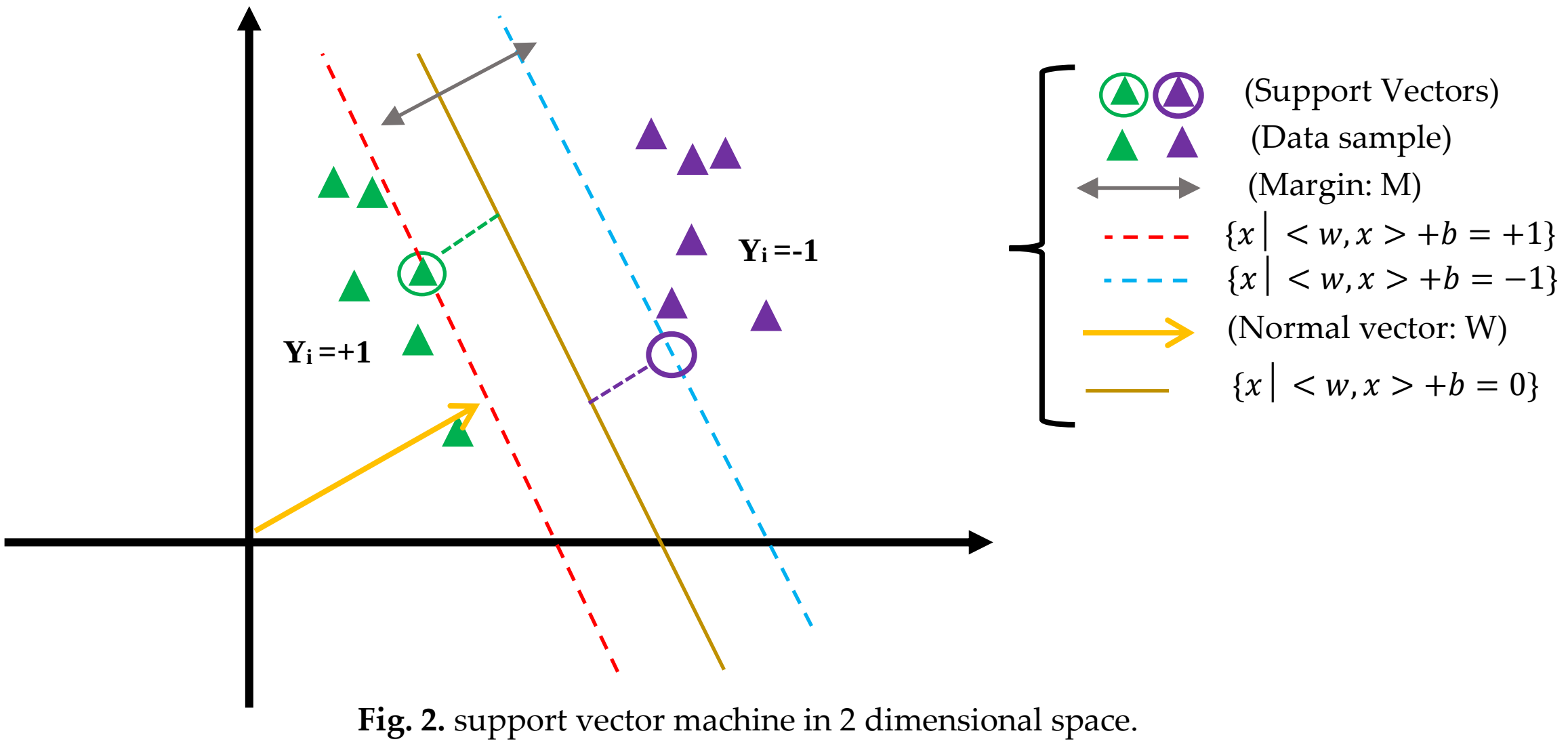

**Fig. 2.** support vector machine in 2 dimensional space.

Given that Fig.2, a description of SVM model is as follows:

Allow training data sample $\{(xi, yi)\}i = 1..n, xi \in R^d$ and the data of the two classes labeled $yi \in \{-1, 1\}$ be separated by a optimal hyperplane in a $\{x|\ < w, x > +b = 0\}$ so that is assigned in the middle of the other two lines, i.e., $\{x|\ < w, x > +b = +1\}$ *and* $\{x|\ < w, x > +b = -1\}$ with margin *M* that the margin $(M = \frac{2}{\|W\|})$ of the separator is the distance between support vectors, data samples closest to the hyperplane are support vectors, and also, b represents the offset between the optimal hyperplane and the origin plane. Then for each training sample $(xi, yi)$:

$$\left\{ \begin{array}{ll} W^T X_i + b \le -\frac{M}{2} & if\ yi = -1 \\ W^T X_i + b \ge \frac{M}{2} & if\ yi = 1 \end{array} \right\} \leftrightarrow Y_i(W^T X_i + b) \ge \frac{M}{2} \quad (14)$$

According to the hyperplane optimization that SVM model was to solve, the optimization problem is as follows [29]:

$$Minimize: \frac{1}{2} \parallel W \parallel^2 \quad subjected\ to: y_i(w.x + b) - 1 \geq 0 \quad \forall i \tag{15}$$

To solve the problem of formula (15), one has to obtain the dual of the problem using the Lagrange Method, namely, *(Lp)*. To obtain the dual form of the problem, the nonnegative Lagrangian coefficients are multiplied by $\alpha$i ≥0. $L_P$ is defined as follows:

$$\mathrm{L_p} = \frac{1}{2} |w|^2 - \sum_i \alpha_i \, (y_i \, (\mathrm{w.x + b}) - 1 \tag{16}$$

Finally, the formula (16) is transformed into the following equation [29]:

$$\text{Maximize} : L_D = \sum_i \alpha_i - \frac{1}{2} \sum_i \sum_j \alpha_i \alpha_j Y_i Y_j \left(X_i X_j\right) \tag{17}$$

Equation (17) is called the dual problem, namely, $L_D$. However, for non-Linear SVM, because there is not the trade-off between maximizing the margin and the misclassification. So, it could not obtain the linear separate hyperplane in over data sample. In the nonlinear space, the best solution, the basic data to higher dimension, i.e., feature space, of linear separate is transformed. At the end, are used the kernel functions, such as linear, polynomial, radial basis function (RBF), and sigmoid [29]. Based on equation (18), $L_D$ for non-Linear data sample is obtained. In (18), parameter *C* is the penalty agent and determines the measure of penalty placed to a fault, so that the "C" value is randomly selected by the user.

$$\text{Maximize } \Phi(\mathrm{W, b}, \xi, \alpha, \beta): L_D = \sum_{i=1}^{N} \alpha_i - \frac{1}{2} \sum_{i=1}^{N} \sum_{j=1}^{N} \alpha_i \alpha_j Y_i Y_j K(X_i, X_j) \tag{18}$$

$$\text{subjected to } \sum_j \alpha_j Y_j = 0, \;\; 0 \leq \alpha_i \leq C, \qquad i = 1, \dots, N$$

N is the number of data sample in (18). In this study, the radial basis function (RBF) [29] is selected as the kernel function as shown in (19):

$$\mathrm{K}\left(X_i, X_j\right) = \exp\, (-\Upsilon \parallel X_i - X_j \parallel^2 \tag{19}$$

In (19), the kernel parameter of $\Upsilon$ with respect to ($\Upsilon \geq 0$) represents the width of the RBF.

### *2.4. Random Trees*

The model of random trees (RTs) is one of the robust predictive models better than other classification models in terms of accuracy computing, data management, more information gain with eliminating fewer features, extract the better rules, working with more data and more complex networks. So, the model for disease diagnosis is suitable. This model consists of multiple trees randomly with high depth so that chooses the most significant votes from a set of possible trees having *K* random features at each node. In other words, in the set of trees, each tree has an equal probability of being assigned. Due to the experiments performed in the classification of the dataset, the accuracy of the RTs model is more accurate than the other models because it uses the evaluation of several features and composes functions. Therefore, RTs can be constructed efficiently and the combination of large datasets of random trees generally leads to proper models. There has been a vast research in the recent years over RTs in the field of machine learning [33]. Generally, Random Trees is confirmed a crucial performance as compared to the classifiers presented as a single tree in this study.

If we consider random trees at very high dimensions with a complex network, then it can include the following steps [33,34]:

1. Using the *N* data sample randomly, in the training dataset to develop the tree.
2. Each node as a predictive feature grasps a random data sample selected so that *m*<*M* (*m* represents the selected feature and *M* represents the full of features in the corresponding dataset. Given that during the growth of trees, m is kept constant.
3. Using the *m* features selected for generating the partition in previous step and computes the *P* node using the best partition path from points. P represents the next node.
4. For aggregating, the prediction dataset uses the tree classification voting from the trained trees with n trees.
5. For generating the terminal RTs model uses the biggest voted features.
6. The RTs process continues until the tree is complete and reaches only one leaf node.

## 3. Related works

In recent years, several studies have been conducted on the diagnose of CAD on different datasets using data mining methods. The most up-to-date dataset that researchers have used recently is the Z-Alizadeh Sani dataset in the field of heart disease. To this end, we review recent research's on the Z-Alizadeh Sani dataset [35,36].

Alizadeh Sani et al. have proposed the use of data mining methods based on ECG symptoms and characteristics in relation to the diagnosis of CAD [37]. In their research, they used Sequential minimal optimization (SMO) and Naïve Bayes algorithms separately and in combination to diagnose the disease. Finally, using the 10-fold cross-validation method for the SMO-Naïve Bayes hybrid algorithm, they achieved more accuracy of 88.52% than the SMO of 86.95% and Naïve Bayes of 87.22% algorithms.

In another study, Alizadeh Sani et al. developed classification algorithms such as SMO, Naïve Bayes, Bagging with SMO and Neural networks for the diagnosis of CAD [12]. Confidence and information gain on CAD have also been used to determine effective features. As a result, among these algorithms, SMO algorithm with information gain has the best performance, with accuracy of 94.08% using 10-fold cross-validation method.

Alizadeh Sani et al. have used computational intelligence methods to diagnose CAD so that they have separately diagnosed three major three coronary stenosis using demographic, symptom and examination, ECG characteristic's, laboratory and echo [38]. They have used analytical methods to investigate the importance of vascular stenosis characteristics. Finally, using the SVM classification model with 10-fold cross-validation method, along with features selection of combined information gain and average information gain, obtained accuracy of 86.14%, 83.17% and 83.50% for left anterior descending (LAD), left circumflex (LCX) and right coronary arteries coronaries (RCA), respectively.

Arabasdi et al., have presented a neural network-genetic hybrid algorithm for the diagnosis of CAD [39]. For this purpose, in their research, genetic and neural network algorithms have been used separately and hybrid to analyze the dataset so that the accuracy of the neural network algorithm and neural network-genetic algorithm using the 10-fold cross-validation method was 84.62% and 93.85%, respectively.

Alizadeh Sani et al. have performed a feature engineering algorithm that have used the Naïve Bayes, C4.5, and SVM classifiers for non-invasive diagnosis of CAD [36]. Given that they have increased their dataset from 303 records to 500 samples. The accuracy obtained using the 10-fold cross-validation method for Naïve Bayes, C4.5, and SVM algorithms were 86%, 89.8%, and 96.40%, respectively.

In a study conducted by Abdar et al. [40], used two-level hybrid genetic algorithm and NuSVM called N2Genetic-NuSVM. Given two-level genetic algorithm, it is used to optimize the SVM parameters and to select the feature in parallel. Using their proposed method, the accuracy of CAD diagnosis was 93.08% through a 10-fold cross-validation method.

## 4. Proposed Methodology

In this section, we follow the proposed methodology in Fig. 3 by IBM Spss Modeler version 18.0 software is used for implementation of classification models.

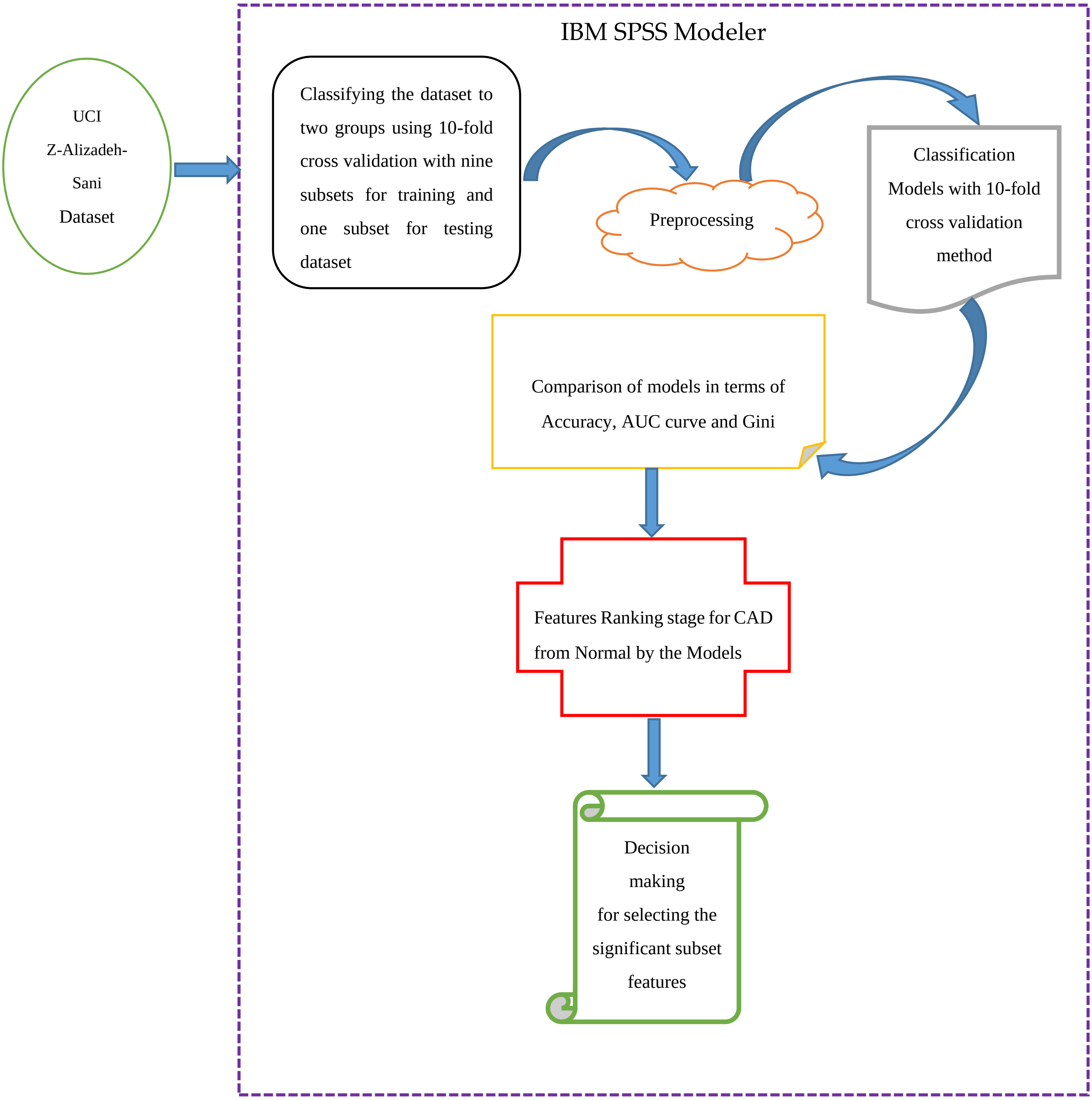

**Fig. 3.** Proposed methodology.

### *4.1. Description of the dataset*

Initially based on Fig. 3, to diagnose the CAD, the Z-Alizadeh Sani dataset is used in this study [35]. This dataset contains information on 303 patients with 55 features, 216 patients with CAD and 88 patients with normal status. The features used in this dataset are divided into 4 groups that are features of CAD for patients including demographics, symptom and examination, electrocardiogram (ECG), and laboratory and echo features described in Table 1. For categorizing the CAD from Normal, the diameter narrowing above 50% is represented a patient as CAD, and its absence is stated as Normal [12].

**Table 1.** Describe the features used in the Z-Alizadeh-Sani dataset with their valid ranges.

| Feature type | Feature name | Range | Measurement | | | |
|---|---|---|---|---|---|---|
| | | | **Mean** | **Std. Error of Mean** | **Std. Deviation** | **Variance** |
| Demographic | Age | [30-80] | 58.90 | 0.6 | 10.39 | 108 |
| Demographic | Weight | [48-120] | 73.83 | 0.69 | 11.99 | 143.7 |
| Demographic | Length | [140-188] | 164.72 | 0.54 | 9.33 | 87.01 |
| Demographic | Sex | Male, Female | --- | --- | --- | --- |
| Demographic | BMI (Body Mass Index Kb/m$^2$) | [18-41] | 27.25 | 0.24 | 4.1 | 16.8 |
| Demographic | DM(Diabetes Mellitus) | [0,1] | 0.3 | 0.03 | 0.46 | 0.21 |
| Demographic | HTN(Hypertension) | [0,1] | 0.6 | 0.03 | 0.49 | 0.24 |
| Demographic | Current smoker | [0,1] | 0.21 | 0.02 | 0.41 | 0.17 |
| Demographic | Ex-smoker | [0,1] | 0.03 | 0.01 | 0.18 | 0.03 |
| Demographic | FH(Family History) | [0,1] | 0.16 | 0.02 | 0.37 | 0.13 |
| Demographic | Obesity | Yes if MBI > 25, No otherwise | --- | --- | --- | --- |
| Demographic | CRF (Chronic Renal Failure) | Yes, No | --- | --- | --- | --- |
| Demographic | CVA(Cerebrovascular Accident) | Yes, No | --- | --- | --- | --- |
| Demographic | Airway disease | Yes, No | --- | --- | --- | --- |
| Demographic | Thyroid disease | Yes, No | --- | --- | --- | --- |
| Demographic | CHF (Congestive Heart Failure) | Yes, No | --- | --- | --- | --- |
| Demographic | DPL(Dyslipidemia) | Yes, No | --- | --- | --- | --- |
| Symptom and examination | BP(Blood Pressure mm Hg) | [90-190] | 129.55 | 1.09 | 18.94 | 358.65 |
| Symptom and examination | PR (Pulse Rate ppm) | [50-110] | 75.14 | 0.51 | 8.91 | 79.42 |
| Symptom and examination | Edema | [0,1] | 0.04 | 0.01 | 0.2 | 0.04 |
| Symptom and examination | Weak peripheral pulse | Yes, No | --- | --- | --- | --- |
| Symptom and examination | Lung rates | Yes, No | --- | --- | --- | --- |
| Symptom and examination | Systolic murmur | Yes, No | --- | --- | --- | --- |
| Symptom and examination | Diastolic murmur | Yes, No | --- | --- | --- | --- |
| Symptom and examination | Typical chest pain | [0,1] | 0.54 | 0.03 | 0.5 | 0.25 |
| Symptom and examination | Dyspnea | Yes, No | --- | --- | --- | --- |
| Symptom and examination | Function class | 1, 2, 3, 4 | 0.66 | 0.06 | 1.03 | 1.07 |
| Symptom and examination | Atypical | Yes, No | --- | --- | --- | --- |
| Symptom and examination | Nonanginal chest pain | Yes, No | --- | --- | --- | --- |
| Symptom and examination | Exertional chest pain | Yes, No | --- | --- | --- | --- |
| Symptom and examination | Low TH Ang (low-Threshold angina) | Yes, No | --- | --- | --- | --- |
| ECG | Rhythm | Sin, AF | --- | --- | --- | --- |
| ECG | Q wave | [0,1] | 0.05 | 0.01 | 0.22 | 0.05 |
| ECG | ST elevation | [0,1] | 0.05 | 0.01 | 0.21 | 0.04 |
| ECG | ST depression | [0,1] | 0.23 | 0.02 | 0.42 | 0.18 |
| ECG | T inversion | [0,1] | 0.3 | 0.03 | 0.46 | 0.21 |
| ECG | LVH (Left Ventricular Hypertrophy ) | Yes, No | --- | --- | --- | --- |
| ECG | Poor R-wave progression | Yes, No | --- | --- | --- | --- |
| Laboratory and echo | FBS (Fasting Blood Sugar mg/dl) | [62-400] | 119.18 | 2.99 | 52.08 | 2712.29 |
| Laboratory and echo | Cr (Creatine mg/dl) | [0.5-2.2] | 1.06 | 0.02 | 0.26 | 0.07 |
| Laboratory and echo | TG (Triglyceride mg/dl) | [37-1050] | 150.34 | 5.63 | 97.96 | 9596.05 |
| Laboratory and echo | LDL (Low-Density Lipoprotein mg/dl) | [18-232] | 104.64 | 2.03 | 35.4 | 1252.93 |
| Laboratory and echo | HDL (High-Density Lipoprotein mg/dl) | [15-111] | 40.23 | 0.61 | 10.56 | 111.49 |
| Laboratory and echo | BUN (Blood Urea Nitrogen mg/dl) | [6-52] | 17.5 | 0.4 | 6.96 | 48.4 |
| Laboratory and echo | ESR (Erythrocyte Sedimentation Rate mm/h) | [1-90] | 19.46 | 0.92 | 15.94 | 253.97 |
| Laboratory and echo | HB (Hemoglobin g/dl) | [8.9-17.6] | 13.15 | 0.09 | 1.61 | 2.59 |
| Laboratory and echo | K (Potassium mEq/lit) | [3.0-6.6] | 4.23 | 0.03 | 0.46 | 0.21 |
| Laboratory and echo | Na(Sodium mEq/lit) | [128-156] | 141 | 0.22 | 3.81 | 14.5 |
| Laboratory and echo | WBC (White Blood Cell cells/ml) | [3700-18.000] | 7562.05 | 138.67 | 2413.74 | 5826137.52 |
| Laboratory and echo | Lymph (Lymphocyte %) | [7-60] | 32.4 | 0.57 | 9.97 | 99.45 |
| Laboratory and echo | Neut (Neutrophil %) | [32-89] | 60.15 | 0.59 | 10.18 | 103.68 |
| Laboratory and echo | PLT (Platelet 1000/ml) | [25-742] | 221.49 | 3.49 | 60.8 | 3696.18 |
| Laboratory and echo | EF (Ejection Fraction %) | [15-60] | 47.23 | 0.51 | 8.93 | 79.7 |
| Laboratory and echo | Region with RWMA | [0-4] | 0.62 | 0.07 | 1.13 | 1.28 |
| Laboratory and echo | VHD (Valvular Heart Disease) | Normal, Mild, Moderate, Severe | --- | --- | --- | --- |
| Categorical | Target Class: Cath | CAD, Normal | --- | --- | --- | --- |

#### *4.2. Classifying the dataset*

Data classification is done into nine subsets i. e., 90% for training the classifiers and one subset i. e., 10% for testing dataset using 10-fold cross-validation.

#### *4.3. Preprocessing the dataset*

Preprocessing step is performed after the data is classified. In general, a set of operations that leading to the creation of a set of cleaned data that can be done on dataset, investigate operation, so-called data preprocessing. The samples values in the Z-Alizadeh-Sani dataset [35] were numeric and string. The purpose of preprocessing the data in this study is to homogenize them so that all data are in the domain of [0 1], which is called the normalization operation, so that is done the standard normalization operation using the Min-Max function. After normalizing numbers, it was time to transform the string data to numeric. In this regard, given the nature of the string data, the value was assigned to them in the interval [0,1]. For example, sex feature has male and female values, which they transformed to zero and one, respectively.

### *4.4. Classifying the models using 10-fold cross-validation method*

For classifying the models was used the 10-fold cross-validation method [41] that the dataset was randomly divided into the same K-scale for the division so and the k-1 subset being used to train the classifiers. The rest of division is also used to investigate the output performance at each step, for 10 times. For this purpose, classifying the prediction models were performed based on the 10-fold cross-validation method so that the average of the criteria was obtained on 10-fold [1,42], which 90% of the data to train and 10% were used for testing the data. Finally, this cross-validation process was executed 10 times so that the results are demonstrated by averaging each ten times.

## 5. Evaluating the results

In this section, we examine the evaluation in two subdivisions. First, evaluation based on the classification criteria, including ROC curve, Gini, Gain, Confidence, Return on investment (ROI), Profit, and Response. Second, evaluation based on significant predictive features.

### *5.1. Evaluation based on classification criteria*

We used a confusion matrix [1, 39, 43,44] to evaluate classification models such as SVM, CHAID, C5.0, and RTs in the diagnosis of CAD on the Z-Alizadeh Sani dataset that described in Table 2.

**Table 2.** Confusion matrix for detection of CAD.

| The Actual class | The predicted class | |
|---|---|---|
| | **Disease (CAD)** | **Healthy (Normal)** |
| Positive | True Positive | False Positive |
| Negative | False Negative | True Negative |

In the following, through the Confusion matrix method, the AUC [1,45] and the Gini index [46] criteria have been obtained, which shown the comparison between the models mentioned for this AUC criterion in Fig. 4 (a,b).

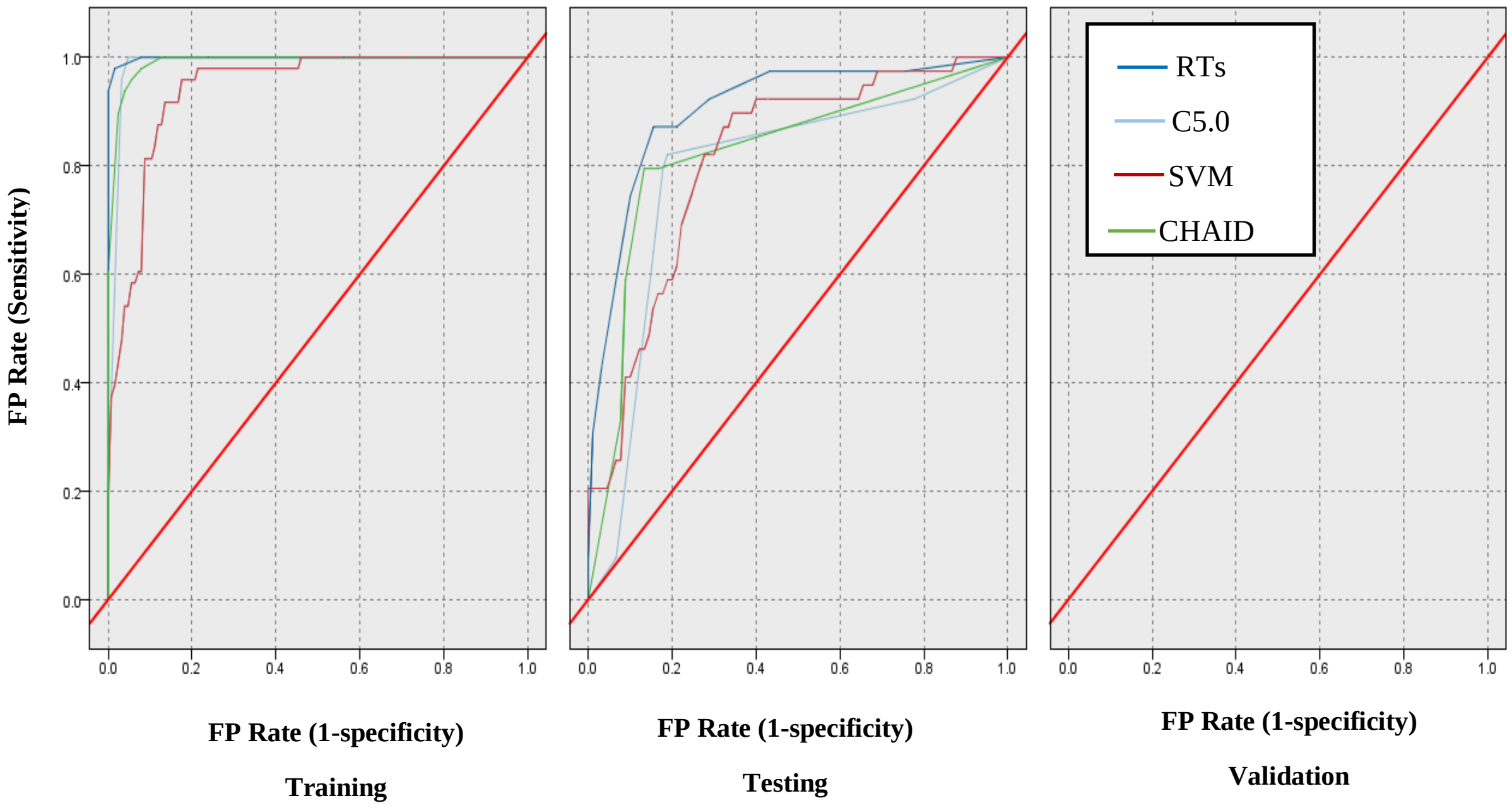

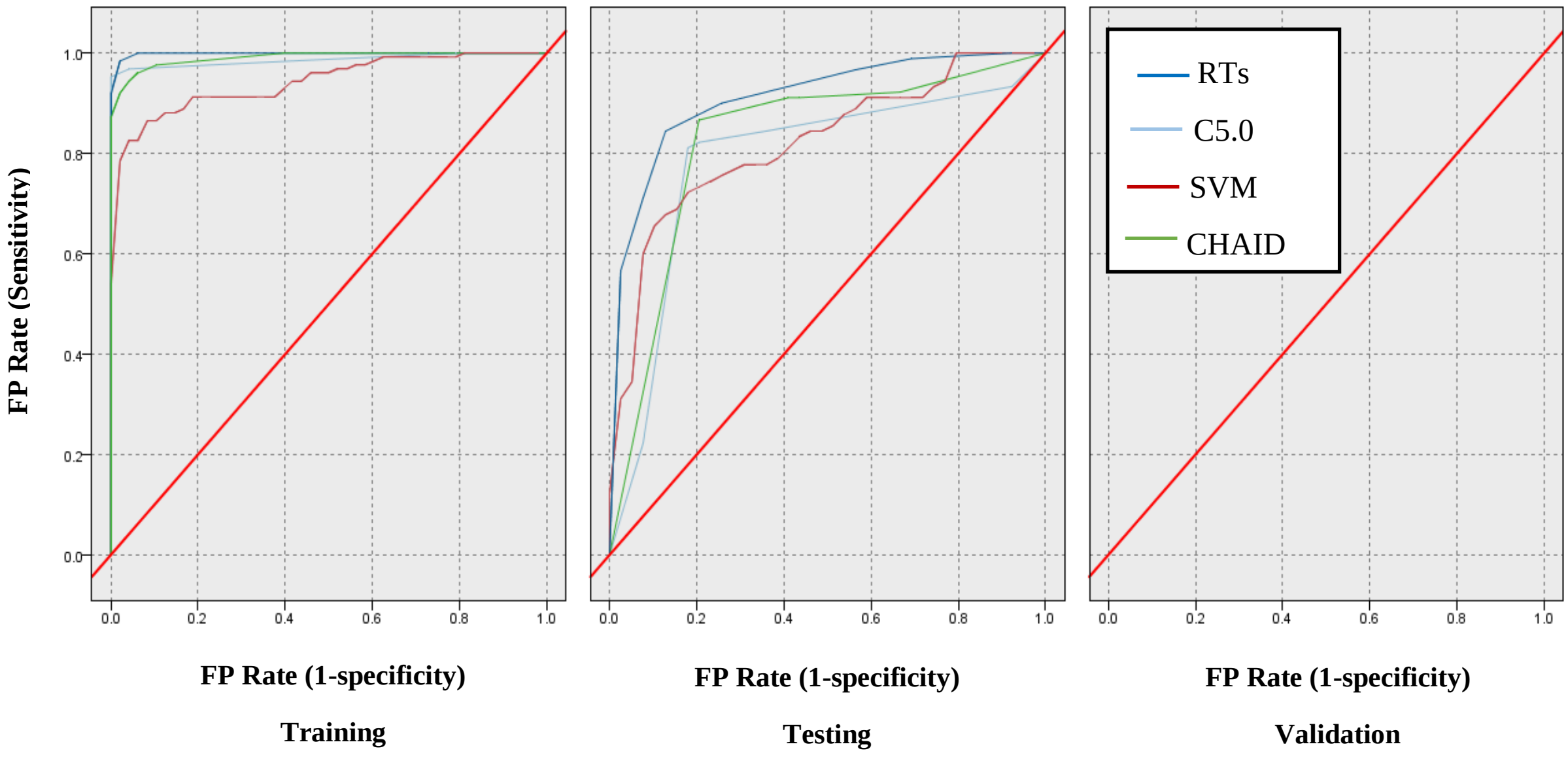

**(b)**

**Fig. 4.** Comparison based on ROC of models: (a) Normal class (b) CAD class.

According to Fig. 4b, the AUC values for the SVM, CHAID, C5.0 and RTs models are 80.90%, 82.30%, 83.00 and 90.50%, respectively. Also the Gini value for SVM, CHAID and RTs models was 61.80%, 64.60%, 66.00% and 93.40%, respectively.

In addition, the Gain, Confidence, Profit, ROI, and Response criteria for evaluating the models have been examined, and comparisons between models through these criteria are shown in Figs 5–9.

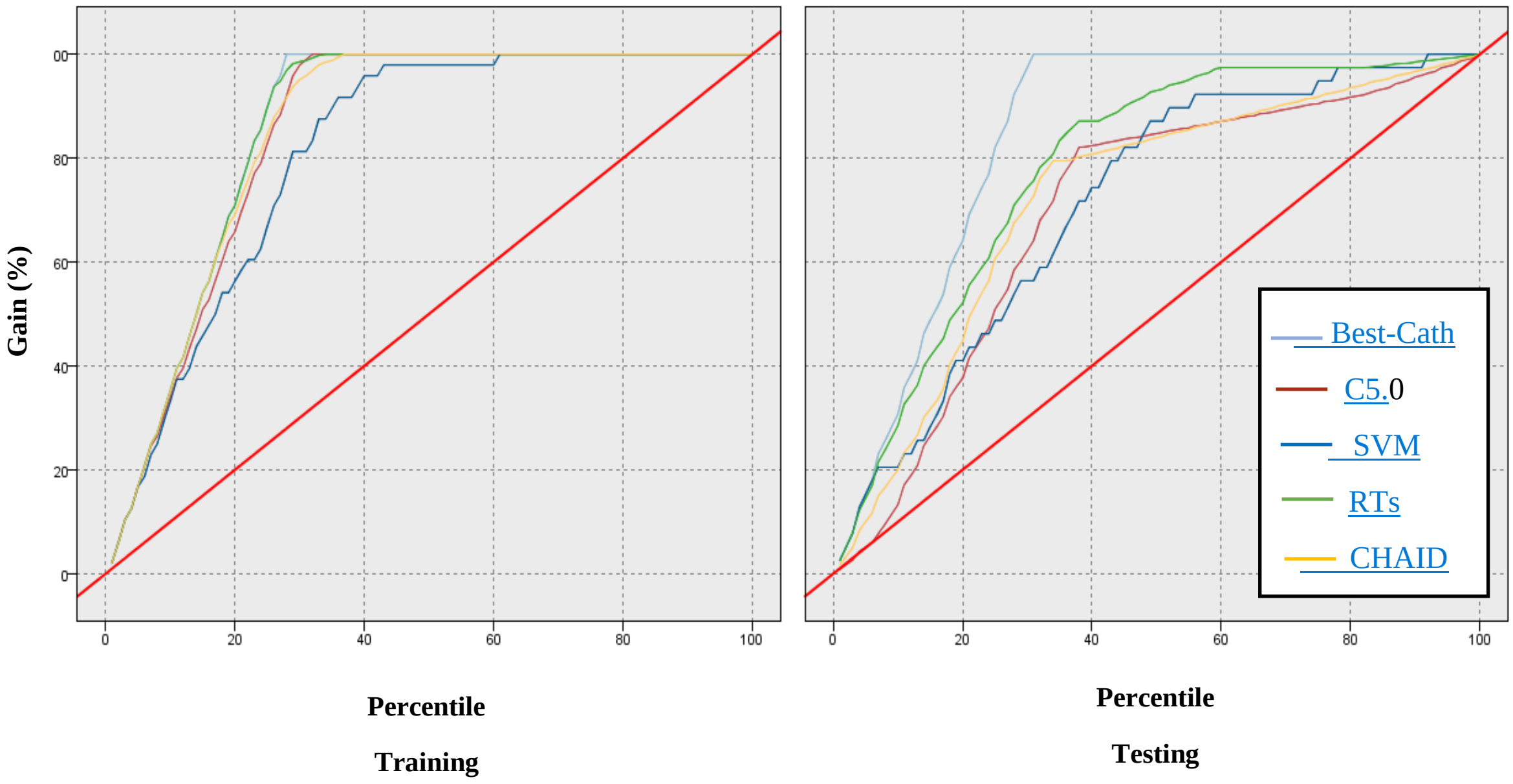

**(a)**

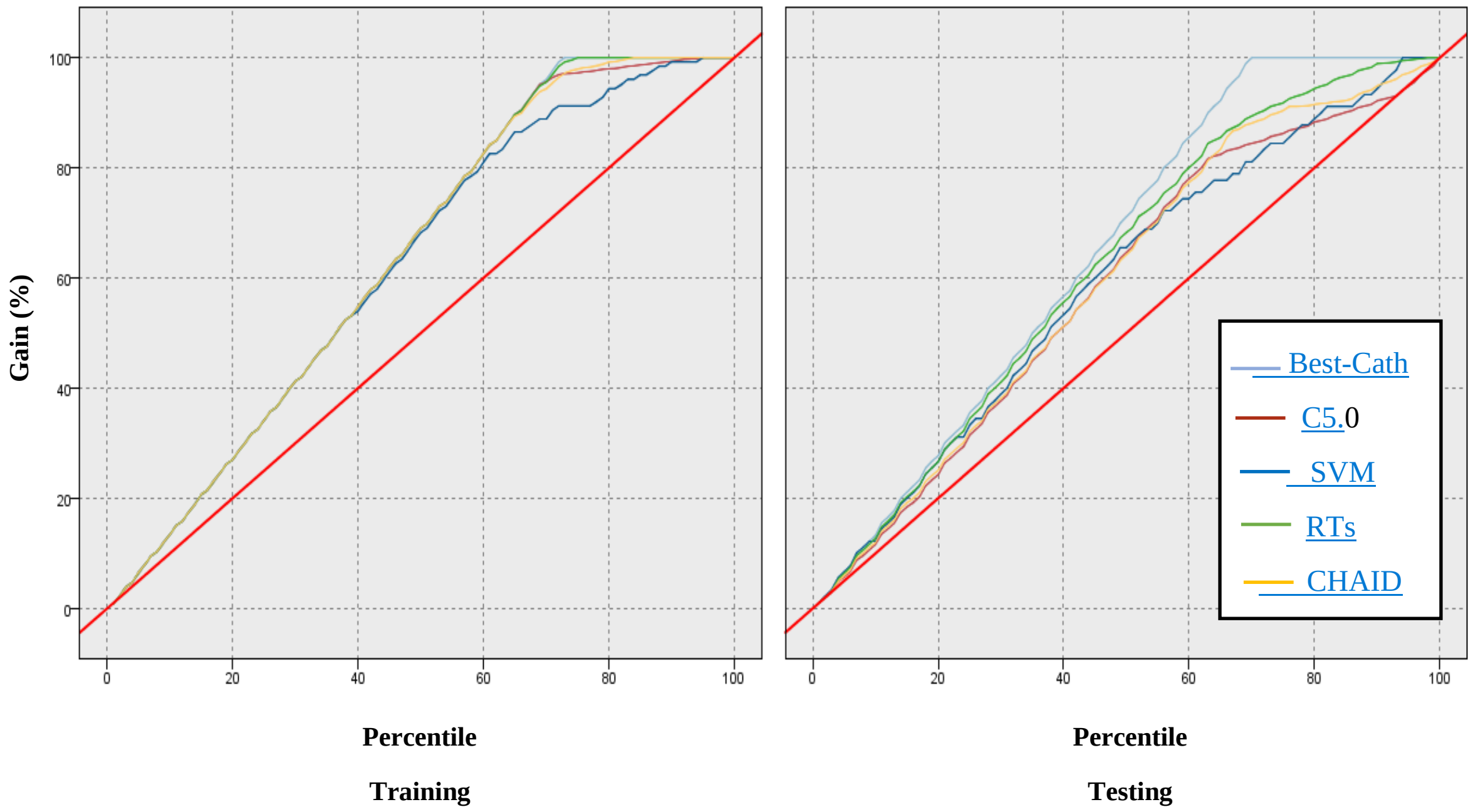

**(b)**

**Fig. 5.** Results based on Gain of models: (a) Normal class, (b) CAD class.

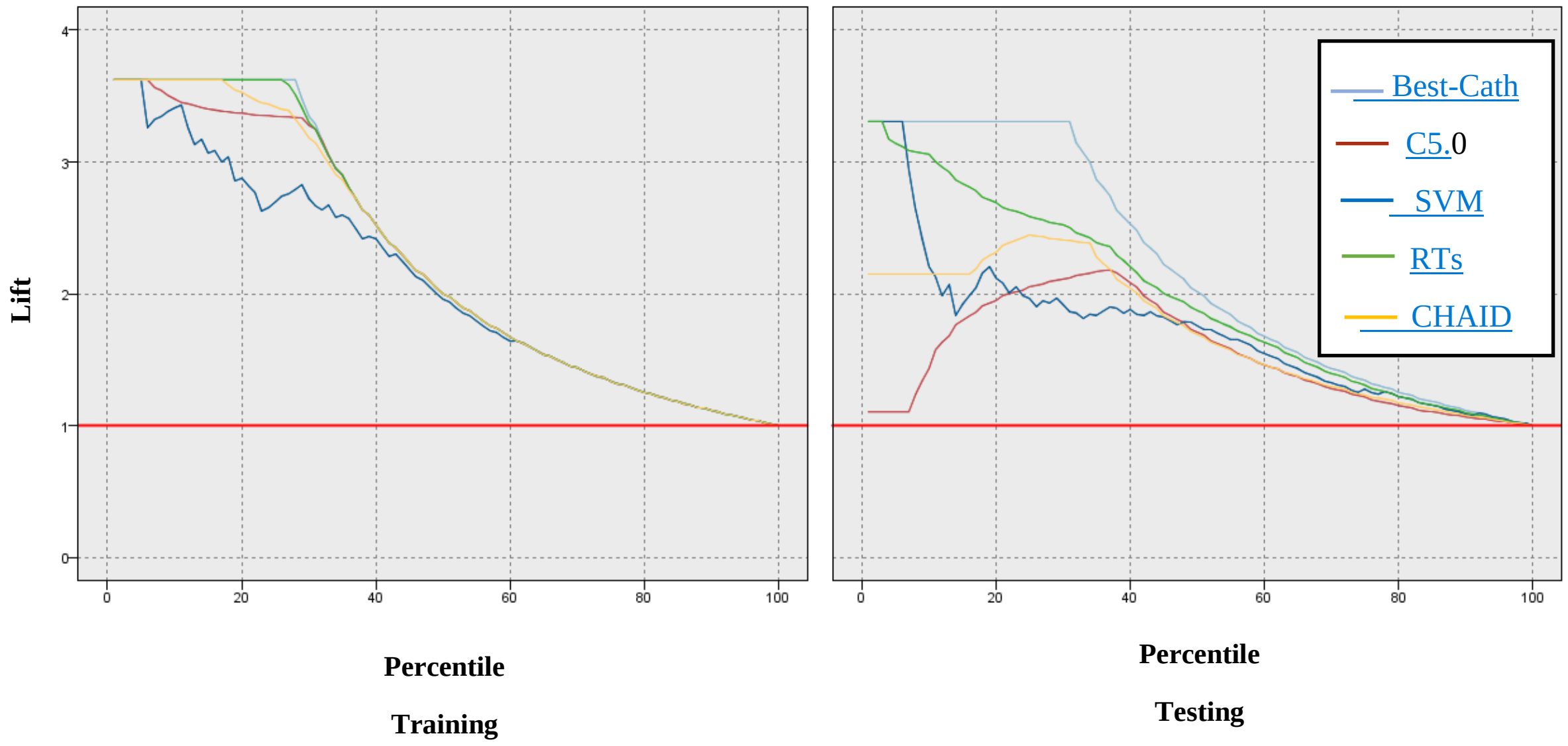

**(a)**

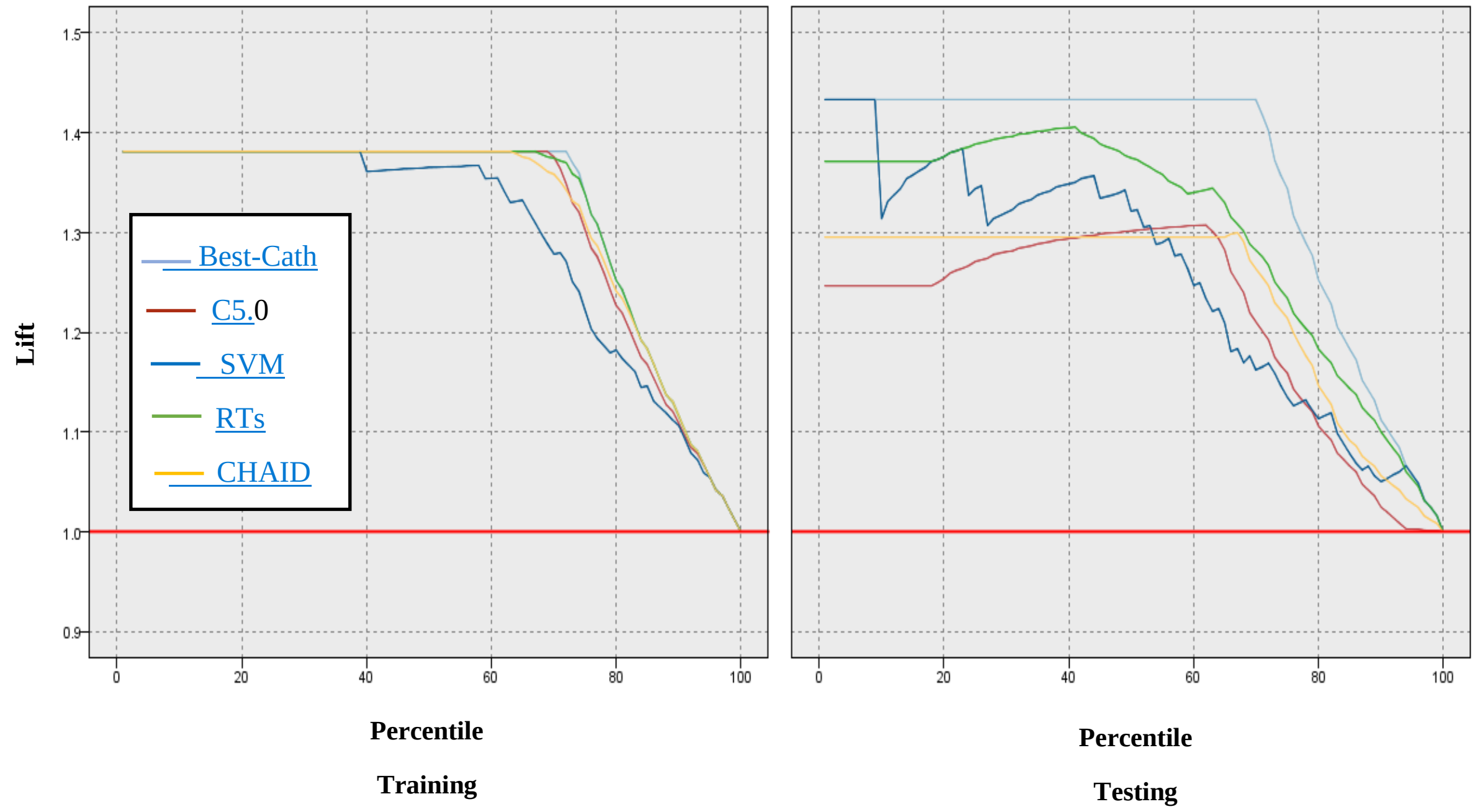

**(b)**

**Fig. 6.** Results based on confidence through the Lift Chart of models: (a) CAD class, (b) Normal class.

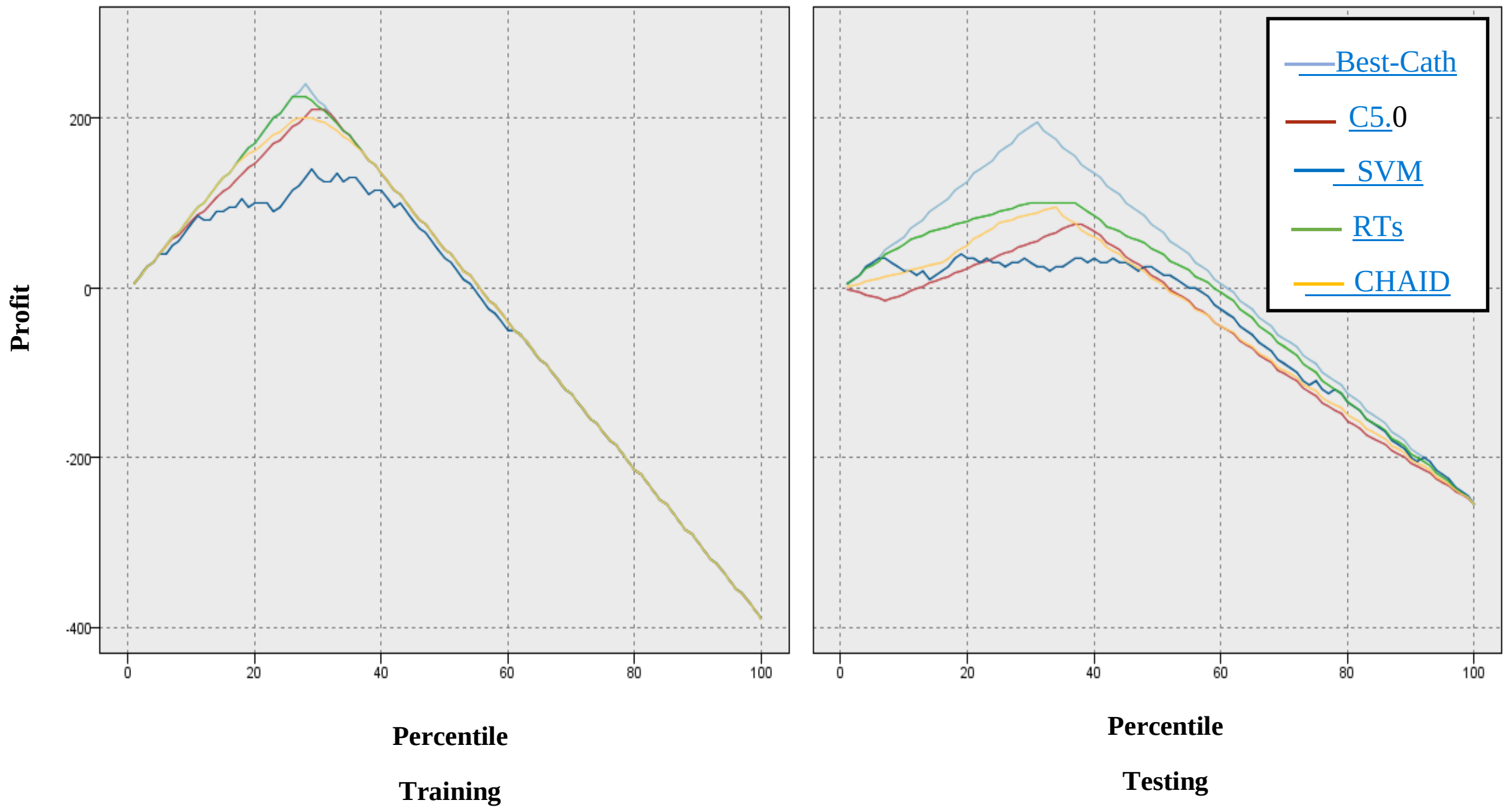

**(a)**

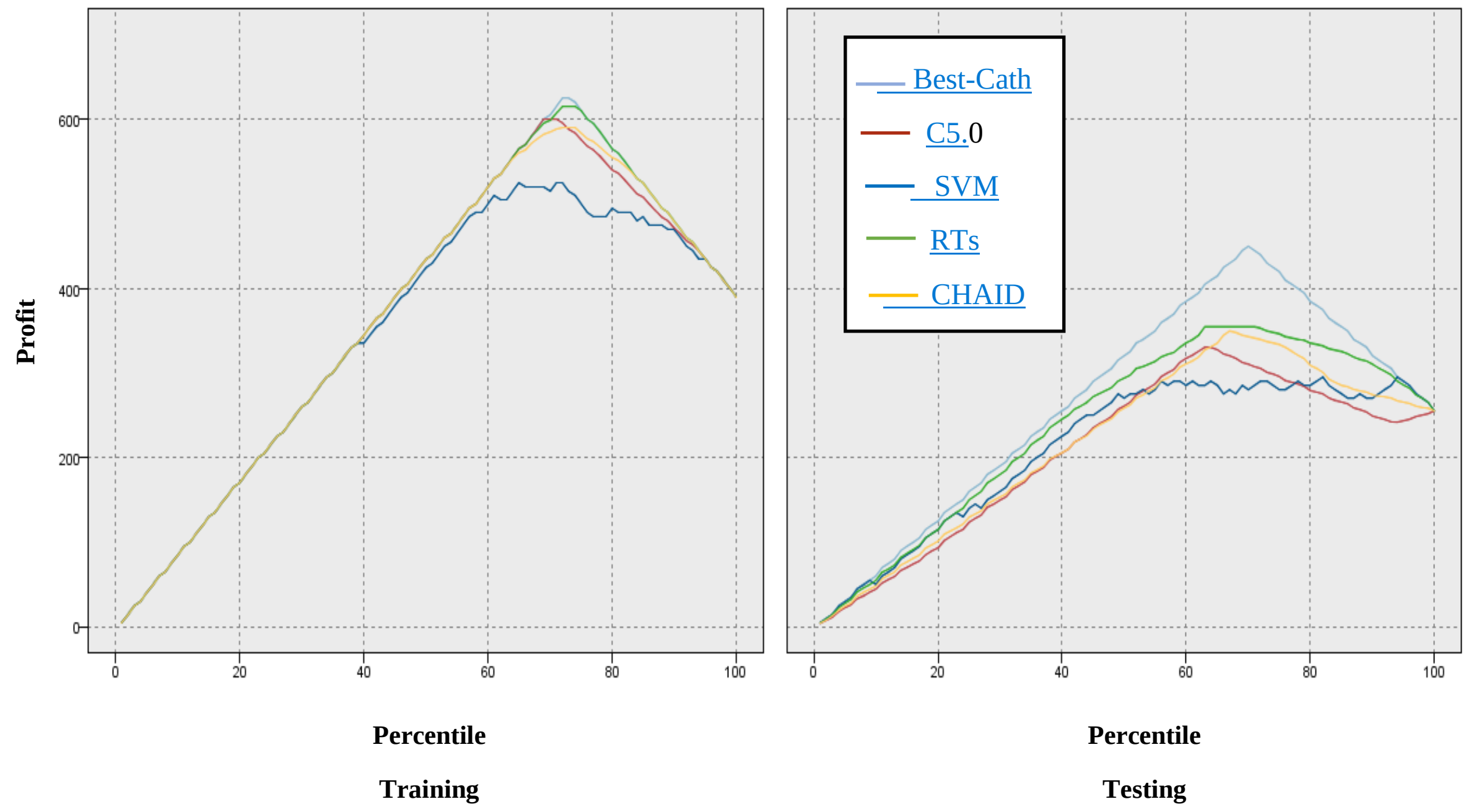

**(b)**

**Fig. 7.** Results based on Profit of models: (a) Normal class, (b) CAD Class.

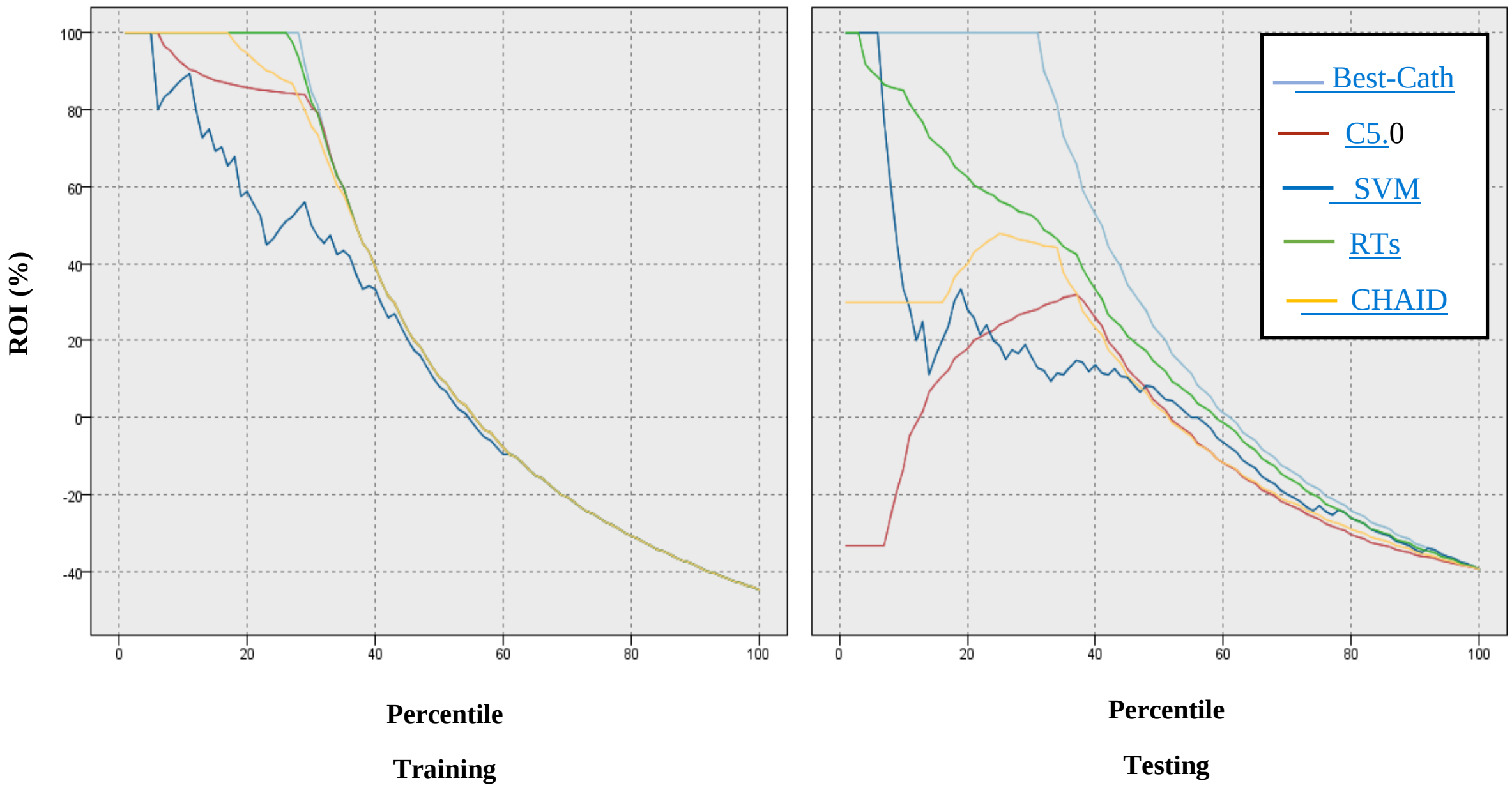

**(a)**

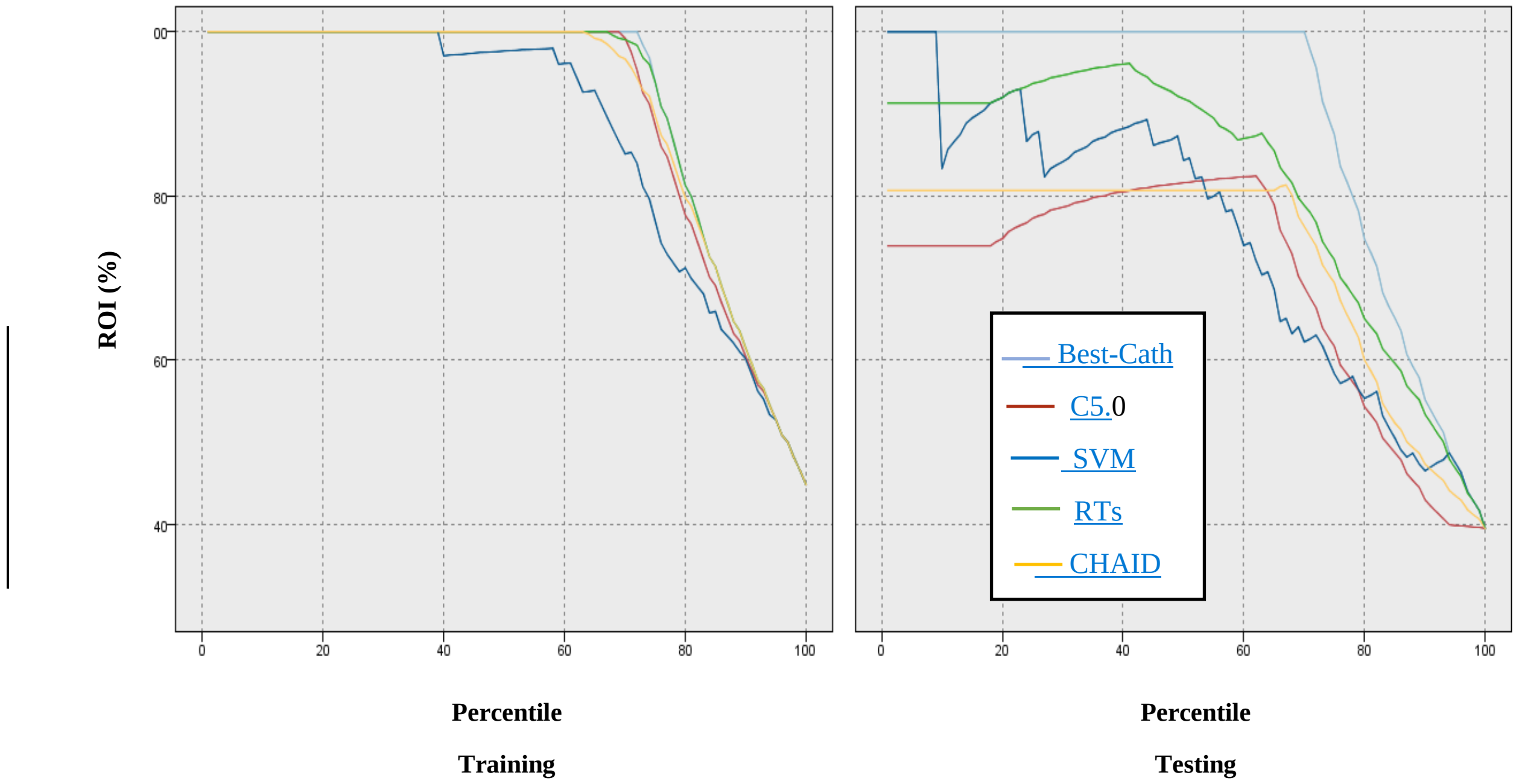

**(b)**

**Fig. 8.** Results based on ROI of models: (a) CAD class, (b) Normal class.

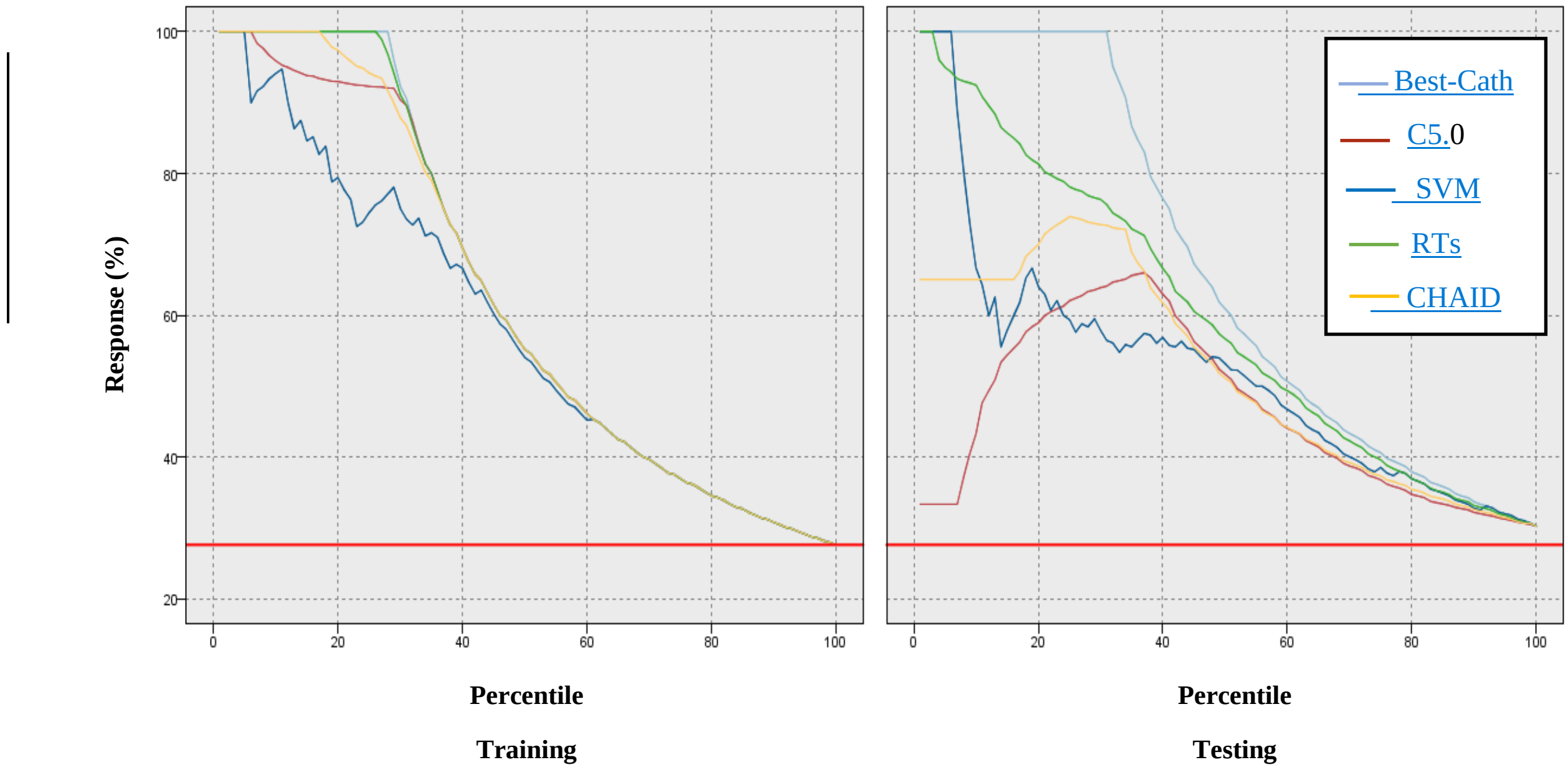

**(a)**

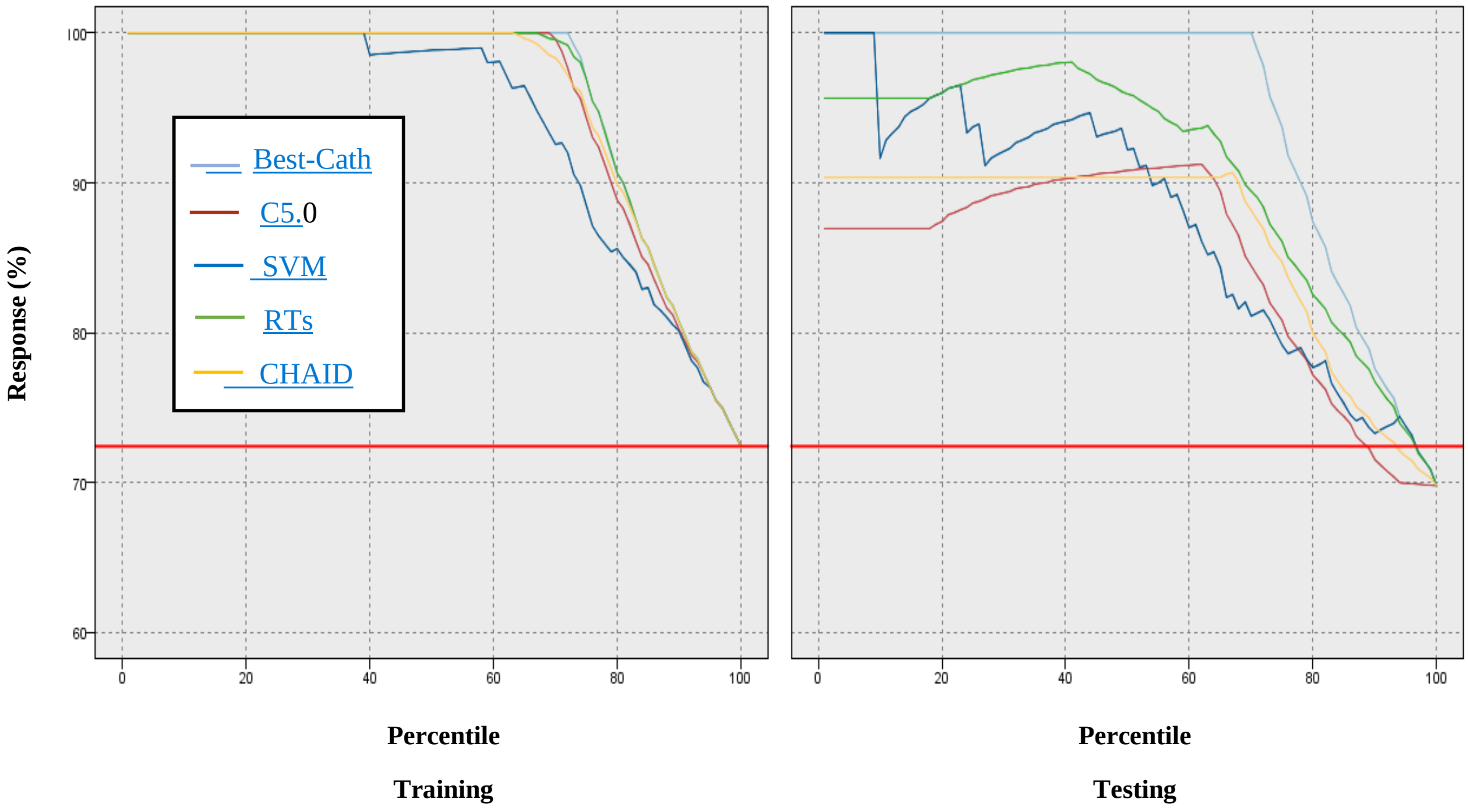

(b)

**Fig. 9.** Results based on Response of models: (a) CAD class, (b) Normal Class.

According to Figs 5–9 of the criteria in the relevant models for the CAD diagnose of Normal class, it can be said that the RTs model has better performance in terms of Gain, Confidence, ROI, Profit and Response criteria than other classification models.

### *5.2. Evaluation based on significant predictive features*

One of the significant evaluates for comparing classification models for predicting the CAD from normal is the use of the importance of predictive features. To this end, we have examined the models in terms of their importance in the ranking stage of features. In fact, the models are measured according to the weight determined to the predictor features. The weighted importance of the features for the models is shown in Tables 2-4.

**Table 2.** Predictor significance imported for features based on ranking for Random trees model.

| No. | Feature | Predictor significance |
|---|---|---|
| 1 | Typical chest pain | 0.98 |
| 2 | TG | 0.66 |
| 3 | BMI | 0.63 |
| 4 | Age | 0.58 |
| 5 | Weight | 0.54 |
| 6 | BP | 0.51 |
| 7 | K | 0.48 |
| 8 | FBS | 0.43 |
| 9 | Length | 0.37 |
| 10 | BUN | 0.3 |
| 11 | PR | 0.29 |
| 12 | HB | 0.26 |
| 13 | Function Class | 0.25 |
| 14 | Neut | 0.25 |
| 15 | EF-TTE | 0.25 |
| 16 | WBC | 0.24 |
| 17 | DM | 0.23 |
| 18 | PLT | 0.2 |
| 19 | Atypical | 0.19 |
| 20 | FH | 0.18 |
| 21 | HDL | 0.16 |
| 22 | ESR | 0.16 |
| 23 | CR | 0.14 |
| 24 | LDL | 0.14 |
| 25 | T inversion | 0.13 |
| 26 | DLP | 0.13 |
| 27 | Region RWMA | 0.12 |
| 28 | HTN | 0.11 |
| 29 | Obesity | 0.1 |
| 30 | Systolic Murmur | 0.09 |
| 31 | Sex | 0.09 |
| 32 | Dyspnea | 0.08 |
| 33 | Current Smoker | 0.06 |
| 34 | BBB | 0.05 |
| 35 | LVH | 0.03 |
| 36 | Edema | 0.02 |
| 37 | EX-Smoker | 0.02 |
| 38 | VHD | 0.01 |
| 39 | St Depression | 0.01 |
| 40 | Lymph | 0.0 |

**Table 3.** Predictor significance imported for features based on ranking for SVM model.

| No. | Feature | Predictor significance |
|---|---|---|
| 1 | Typical chest pain | 0.04 |
| 2 | Atypical | 0.03 |
| 3 | Sex | 0.02 |
| 4 | Obesity | 0.02 |
| 5 | FH | 0.02 |
| 6 | Age | 0.02 |
| 7 | DM | 0.02 |
| 8 | Dyspnea | 0.02 |
| 9 | Systolic Murmur | 0.02 |
| 10 | St Depression | 0.02 |
| 11 | HTN | 0.02 |
| 12 | LDL | 0.02 |
| 13 | Current Smoker | 0.02 |
| 14 | DLP | 0.02 |
| 15 | BP | 0.02 |
| 16 | LVH | 0.02 |
| 17 | Nonanginal | 0.02 |
| 18 | Tin version | 0.02 |
| 19 | Length | 0.02 |
| 20 | Function Class | 0.02 |
| 21 | BBB | 0.02 |
| 22 | VHD | 0.02 |
| 23 | CHF | 0.02 |
| 24 | PR | 0.02 |
| 25 | WBC | 0.02 |
| 26 | BUN | 0.02 |
| 27 | FBS | 0.02 |
| 28 | ESR | 0.02 |
| 29 | CVA | 0.02 |
| 30 | Thyroid Disease | 0.02 |
| 31 | Lymph | 0.02 |
| 32 | Weight | 0.02 |
| 33 | CR | 0.02 |
| 34 | Airway disease | 0.02 |
| 35 | TG | 0.02 |
| 36 | CRF | 0.02 |
| 37 | Diastolic Murmur | 0.02 |
| 38 | Low TH Ang | 0.02 |
| 39 | Exertional CP | 0.02 |
| 40 | Weak Peripheral Pulse | 0.02 |
| 41 | Neut | 0.02 |
| 42 | PLT | 0.02 |
| 43 | St Elevation | 0.02 |
| 44 | EF-TTE | 0.02 |
| 45 | K | 0.02 |
| 46 | BMI | 0.02 |
| 47 | EX-Smoker | 0.02 |
| 48 | Lung rales | 0.02 |
| 49 | HDL | 0.02 |
| 50 | Na | 0.01 |
| 51 | Edema | 0.01 |
| 52 | Q Wave | 0.01 |
| 53 | HB | 0.01 |
| 54 | Poor R Progression | 0.01 |
| 55 | Region RWMA | 0.01 |

**Table 4.** Predictor significance imported for features based on ranking for C5.0 model.

| No. | Feature | Predictor significance |
|---|---|---|
| 1 | Typical chest pain | 0.28 |
| 2 | CR | 0.14 |
| 3 | ESR | 0.13 |
| 4 | T inversion | 0.1 |
| 5 | Edema | 0.09 |
| 6 | Region RWMA | 0.08 |
| 7 | Poor R Progression | 0.04 |
| 8 | Sex | 0.03 |
| 9 | DM | 0.03 |
| 10 | BMI | 0.02 |
| 11 | WBC | 0.02 |
| 12 | DLP | 0.02 |
| 13 | Length | 0.01 |
| 14 | Dyspnea | 0.0 |
| 15 | EF-TTE | 0.0 |

**Table 5.** Predictor significance imported for features based on ranking for CHAID model.

| No. | Feature | Predictor significance |
|---|---|---|
| 1 | Typical chest pain | 0.33 |
| 2 | Age | 0.15 |
| 3 | T inversion | 0.11 |
| 4 | VHD | 0.1 |
| 5 | DM | 0.09 |
| 6 | HTN | 0.04 |
| 7 | Nonanginal | 0.03 |
| 8 | BP | 0.02 |
| 9 | Region RWMA | 0.02 |
| 10 | HDL | 0.02 |

## 5. Results and discussion

In the modeling process proposed in Section 4, we implemented several data mining models including SVM, CHAID, C5.0, RTs. The 10-fold cross-validation method was used to build these models so that the data was divided into train (90 percent) and test (10 percent) subsets. The results show that the Random trees model is the best classification model compared to the other models so that the accuracy of the RTs model is obtained 91.47% using 10-fold cross-validation method. While the accuracy of SVM, CHAID, and C5.0 models were 69.77%, 80.62%, and 82.17%, respectively.

Given that the accuracy is computed using the following formula (TP+TN/(TP+TN+FP+FN)) where TP is True Positive, TN is True Negative, FP is False Positive, and FN is False Negative [1].

Another criterion for evaluating the models in this study was the AUC criterion, which is obtained 80.90%, 82.30%, 83.00%, and 96.70 for SVM, CHAID, C5.0, and RTs, respectively.

Furthermore, achievement of this study is the use of criteria that were not found in previous studies, including Gain, Confidence, ROI, Profit, and Response, as shown in Figs 5-9, in terms of these criteria, the Random trees model has the best performance, other classification models.

Finally, based on Tables 2 to 5, it can be resulted that in each of the 4 models the Typical chest pain feature is selected as the most significant predictor so that the predictor significance of the Typical chest pain feature for the random trees model is equal to 0.98 with the most significant and the least significant for Lymph feature is equal to zero. Given that for the features, intervals 1 and 2 are applied in the simulator. In Table 1, also the Typical chest pain is the most significant feature with a value of 0.04 and the Region RWMA as the least significant feature of 0.01 was obtained. According to Tables 4 and 5 Typical chest pain as the most significant feature is equal to 0.28 and 0.33 respectively, and the least significant feature according to Table 4 for EF-TTE feature equal to zero and the least significant feature according to Table 5 is 0.02. It is therefore confirmed that the RTs model is the best model relative to other classification models according to the above tables.

One of the advantages of the Random Trees model were the most significant obtained rules of CAD diagnosis that is placed in Table 6.

**Table 6.** the most significant obtained rules for CAD diagnosis using Random trees (Top Decision Rules for 'Cath' class.

| Decision Rule | Most Frequent Category | Rule Accuracy | Forest Accuracy | Interestingness Index |
|---|---|---|---|---|
| (BP>110.0), (FH>0.0), (Neut>51.0) and (Typical Chest Pain>0.0) | CAD | 1.000 | 1.000 | 1.000 |
| (BMI<=29.02), (EF-TTE>50.0), (CR<= 0.9), (Typical Chest Pain > 0.0) and (Atypical={N}) | CAD | 1.000 | 1.000 | 1.000 |
| (Wight > 8.0), (CR > 0.9), (Typical Chest Pain > 0.0) and (Atypical= {N}) | CAD | 1.000 | 1.000 | 1.000 |
| (K<= 4.9), (WBC>5700.0), (CR< 0.9), | CAD | 1.000 | 1.000 | 1.000 |
| (DM >0.0) and (Typical Chest Pain > 0.0) | CAD | 1.000 | 1.000 | 1.000 |

According to Table 6, the extracted rules for CAD are described as follows:

If the condition is true of (BP> 110.0), (FH> 0.0), (Neut> 51.0) and (Typical Chest Pain> 0.0), then the CAD exist highly accurate and also interestingness index, otherwise, the person is normal. If the condition is true of (Typical Chest Pain > 0.0) and (Atypical={N}), then like the result it is like the result of previous conditions. In the following, if the condition is true of (Wight > 8.0), (CR > 0.9), (Typical Chest Pain > 0.0) and (Atypical= {N}), then the person is abnormal or CAD. Finally, if (K<= 4.9), (WBC>5700.0), (CR< 0.9), (DM >0.0) and (Typical Chest Pain > 0.0) is true, as a result the person is abnormal.

In recent years, several studies have been conducted on the diagnosis of CAD on different datasets using data mining methods. To this end, we review recent research on the updated Z-Alizadeh Sani dataset that these are described in Table 7. Given that the results of the Accuracy, AUC and Gini criteria for the models have been done according to the 10-fold cross validation method compared to previous studies.

**Table 7.** The performed works for CAD diagnosis on the Z-Alizadeh Sani dataset with 10-fold cross validation method.

| Referense | Methods | No. features subset selection | Accuracy (%) | Auc % | Gini % |
|---|---|---|---|---|---|
| [37] | Naïve Bayes-SMO | 16 | 88.52 | Not reported | Not reported |
| [12] | SMO along with information Gain | 34 | 94.08 | Not reported | Not reported |
| [38] | SVM along with average information gain and also information gain | 24 | 86.14 for LAD<br>83.17 for LCX<br>83.50 for RCA | Not reported | Not reported |
| [39] | Neural Network-Genetic Algorithm-weight by SVM | 22 | 93.85 | Not reported | Not reported |
| [36] | SVM along with Feature engineering | 32 | 96.40 | 92 | Not reported |
| [40] | N2Genetic-nuSVM | 29 | 93.08 | Not reported | Not reported |
| In our study | Random Trees | 40 | 91.47 | 96.70 | 93.40 |

Taking a look at Table 7, it can be resulted that the proposed method based on Random Trees outperforms other methods in terms of accuracy, AUC, and Gini criteria. It implies that the 40 features extracted by using RTs are the most informative ones about the CAD disease.

## 7. Conclusion and future works

In this study, a computer-aided diagnosis system was used to diagnose CAD as a common heart disease on the Z-Alizadeh Sani dataset [35] so that this system is implemented using the IBM Spss Modeler version 18.0 tool. Since angiography is the most common tool of diagnosis of heart disease, it has a cost and side effects for individuals. So artificial intelligence methods, that is, machine learning techniques can be a solution to the stated challenge. Hence, we have such classification models including SVM, CHAID, C5.0, and Random trees are used for modeling with 10-fold cross-validation method, which are based on accuracy, AUC, Gini, ROI, Profit, Confidence, response, and Gain have been examined and evaluated. Finally, based on the criteria stated, the Random trees model is found as the best model than the other models so that select the predictive features based on the order of their priority with the highest accuracy, we conclude that the Random trees model with the most significant features of 40 and the accuracy of 91.47% has better performance than the other classification models. So as to, with this number of features, we will have more information gain than the features selected in previous works. Another achievement of this study was the important extraction rules for CAD diagnosis using the Random Trees model that these rules are shown in Table 6. As future work, the fuzzy intelligent system can be used in combination with artificial intelligence models to diagnose CAD on the Z-Alizadeh Sani dataset and other datasets. Another way to better diagnose CAD disease on this dataset and other real datasets is deep learning models and combining deep learning approaches with a distributed design and architecture.

**Acknowledgments:** We acknowledge the financial support of this work by the Hungarian State and the European Union under the EFOP-3.6.1-16-2016-00010 project.

**Conflicts of Interest:** The authors declare no conflict of interest.

**Contributions:** Conceptualization, Seyyed Mohammad Razavi and Amir Mosavi; Data curation, Javad Hassannataj Joloudari, Edris Hassannataj Joloudari, Mohammad Ghasemigol, and Narjes Nabipour; Formal analysis, Javad Hassannataj Joloudari, Mohammad Ghasemigol and Narjes Nabipour; Funding acquisition, Shahaboddin Shamshirband; Investigation, Edris Hassannataj Joloudari, Hamid Saadatfar, Mohammad Ghasemigol, and Seyyed Mohammad Razavi; Methodology, Edris Hassannataj Joloudari, Amir Mosavi, and Shahaboddin Shamshirband; Project administration, Hamid Saadatfar; Resources, Javad Hassannataj Joloudari, Hamid Saadatfar, and Seyyed Mohammad Razavi; Software, Edris Hassannataj Joloudari, and Hamid Saadatfar; Supervision, Mohammad Ghasemigol, and Laszlo Nadai; Validation, Narjes Nabipour; Visualization, Mohammad Ghasemigol, and Seyyed Mohammad Razavi; Writing – original draft, Javad Hassannataj Joloudari, and Hamid Saadatfar; Writing – review & editing, Javad Hassannataj Joloudari, Laszlo Nadai, Mohammad Ghasemigol, and Amir Mosavi; checking the results, Laszlo Nadai.